# The Impact of Partner Expressions on Felt Emotion in the Iterated Prisoner's Dilemma: An Event-level Analysis

**Maria Angelika-Nikita**  MARIAANG@STANFORD.EDU
Stanford University

**Celso M. de Melo**  CELSO.MIGUEL.DE.MELO@GMAIL.COM
DEVCOM Army Research Laboratory

**Kazunori Terada**  TERADA@GIFU-U.AC.JP
Gifu University

**Gale Lucas**  LUCAS@ICT.USC.EDU
**Jonathan Gratch**  GRATCH@ICT.USC.EDU
University of Southern California

## Abstract

Social games like the prisoner's dilemma are often used to develop models of the role of emotion in social decision-making. Here we examine an understudied aspect of emotion in such games: how an individual's feelings are shaped by their partner's expressions. Prior research has tended to focus on other aspects of emotion. Research on felt-emotion has focused on how an individual's feelings shape how they treat their partner, or whether these feelings are authentically expressed. Research on expressed-emotion has focused on how an individual's decisions are shaped by their partner's expressions, without regard for whether these expressions actually evoke feelings. Here, we use computer-generated characters to examine how an individual's moment-to-moment feelings are shaped by (1) how they are treated by their partner and (2) what their partner expresses during this treatment. Surprisingly, we find that partner expressions are far more important than actions in determining self-reported feelings. In other words, our partner can behave in a selfish and exploitive way, but if they show a collaborative pattern of expressions, we will feel greater pleasure collaborating with them. These results also emphasize the importance of context in determining how someone will feel in response to an expression (i.e., knowing a partner is happy is insufficient; we must know what they are happy-at). We discuss the implications of this work for cognitive-system design, emotion theory, and methodological practice in affective computing.

## 1. Introduction

Emotions are social (Manstead, Fischer, & Jakobs, 1999). Of course, people take pleasure in maximizing self-interest, but they feel joy in contributing to the common good. In competitive settings, people even take pleasure in causing their opponent harm. These social feelings can shape decisions, even in interactions with machines (Breazeal & Brooks, 2005; Giannopulu, Terada, &

Watanabe, 2018). Indeed, emotional feelings are central to many theories of why people depart from rational choice theory (Camerer, 2003; Fehr & Schmidt, 1999; Frank, 1988; Loewenstein & Lerner, 2003). These theories argue that people take pleasure in selfless acts, take offence when others act selfishly, and feel guilt at their own selfish acts. Adding such feelings into economic models can better explain why people split money near-equally in the dictator game and cooperate in the prisoner's dilemma when game theory predicts they should not.

The balance between selfish or selfless feelings can vary by personality and culture (Hofstede, 2011; Murphy, Ackermann, & Handgraaf, 2011), but here we provide evidence that this balance is also shaped by the emotional expressions of others. When faced with a partner that displays "competitive expressions" (i.e., emotional expressions that suggest the partner prioritizes their own self-interests), people report experiencing stronger selfish emotions. When faced with a partner that displays "cooperative expressions" (i.e., emotional expressions that suggest the partner prioritizes the common good), people report experiencing stronger selfless emotions. These effects cannot be easily explained by emotional contagion; one common theory of how people "catch" emotion from the expressions of others (Hatfield, Cacioppo, & Rapson, 1994). Nor can they be explained by "affect as information" theories that argue that the expressions of others serve as information that informs decisions (van Kleef, 2008). Rather, this finding suggests that people internalize the *expressed* values of their partners, and these are reflected in the emotions they experience. Most surprisingly, and the most novel contribution of the present study, we show this effect is not determined by the partner's *actions* (i.e., does the partner act in a selfish or selfless way), but primarily by the partner's emotional *expressions*. In other words, our partner can behave in a selfish and exploitive way, but if they show a collaborative pattern of expressions, we will feel greater pleasure collaborating with them, and diminished pleasure at maximizing self-interest.

In addition to highlighting a novel relationship between expression and emotion, our results reinforce recent theoretical and methodological advances in the field of affective computing. First, our findings underscore an important distinction between emotion and *emotion-AT*. Research in affective computing often focuses on recognizing emotion (i.e., does a person feel joy?). But emotions have intentionality (Goldie, 2002), and our results emphasize the importance of identifying the target of the emotion (i.e., do they feel *joy-at* being selfish, or *joy-at* being selfless). The intentionality of emotion also has implications for unit of analysis in studying affective signals. In that emotions (in contrast to moods) are balanced responses to specific events, they should be studied at the level of emotion-relevant events. Often, affective responses are averaged over an entire interaction which can obscure important patterns concerning the target of these feelings. Finally, the current study highlights the importance of virtual agents as methodological tools to uncover affective processes. Agents allow rich social interactions while maintaining a high degree of experimental control (Blascovich et al., 2002; de Melo, Carnevale, & Gratch, 2014).

Here we examine the impact of emotion expression on behavior and feelings in the context of a 20-round iterated prisoner's dilemma task (a standard social dilemma used to study the impact of emotion on decision-making). In this IRB-approved study, participants played for monetary reward and were told they were playing another participant online. In fact, they played a computer agent that varied in its behavior and emotional expressions. The agent made decisions following either an "extortionist" or "generosity" strategy across the 20 rounds (Press & Dyson, 2012; Stewart & Plotkin, 2013). These are contingent-strategies proposed in game theory to maximize selfish or collective reward, respectively. Expressively, agents displayed "competitive" or "cooperative" pattern of emotional reactions to game events, following an "expression policy" previously shown to convey selfish or selfless intentions (de Melo, Carnevale, Read, & Gratch, 2014). After each round in the iterated game, participants were asked to report how they felt about the joint decision. Overall, we find that the previously-observed pattern of opponent expressions has a strong effect on participant self-reported *feelings*, whereas the opponent's preceding pattern of decisions has a strong effect over participant *decisions*. Together, these findings suggest a looser link between feelings and actions than some decision-making models might suggest.

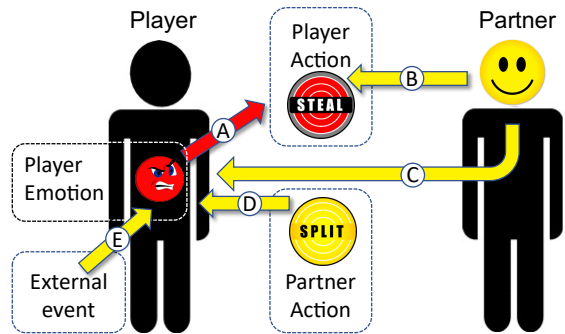

Fig. 1: The pathways by which emotion influences player feelings and decisions in social games like the prisoner's dilemma: (A) player feelings shape player decisions; (B) partner expressions shape player decisions; (C) partner expressions shape player feelings; (D) partner actions shape player feelings; and (E) exogenous event shapes player feelings. The present paper focuses on the (C) and (D) paths.

The next section reviews the prior research on emotion and social decision-making. First, we highlight the various pathways by which expression, feeling, and action interact and shape subsequent behavior and feelings. We next review the unit of analysis typically used to study emotion and feeling, arguing for the importance of focusing on individual outcomes rather than aggregate measures that treat the entire game as a unit of analysis. Section 3 presents our novel analysis of a large existing corpus of player behavior in the iterated prisoner's dilemma that we have developed within our group. Finally, we discuss important limitations of this analysis, qualifications on our findings, and next steps.

## 2. Feeling, Expression, and Action

For the context of this paper, we use "feeling" to refer to self-reported experience of emotion, which has a debatable connection to physiological processes or facial expressions (Fridlund, 1997; Scarantino, 2017). A large body of research has explored the role of felt emotion in decision-making. In individual decision-making, felt emotions are implicated in why people depart from the predictions of decision theory (Loewenstein & Lerner, 2003; Mellers, Schwartz, Ho, & Ritov, 1997). When making decisions with other people, both felt and expressed emotions are claimed to explain why people depart from game-theoretic predictions (Camerer, 2003; Frank, 1988).

Research often uses classic game-theoretic tasks, such as the prisoner's dilemma to highlight emotion's influence and we follow this tradition. Besides informing social theories, such dilemmas inform human-machine interaction including decision with robots (Correia et al., 2019) and self-driving cars (Gogoll & Müller, 2017). In the prisoner's dilemma game, two players make a simultaneous decision to either cooperate or defect. They receive a payout, known in advance, based on this joint decision. If they both cooperate, they each receive a reward $R$ for cooperating. If they both defect, they receive a punishment payoff $P$ that is lower than $R$. However, if one cooperates and the other defects, the defector (i.e., the exploiting party) earns a high temptation reward $T$ and the cooperator (i.e., the exploited party) receives a "sucker's" payout $S$. In the classic formulation, $T > R > P > S$. According to rational choice theory (which assumes players only consider their individual payout), the rational decision is for both players to defect. The iterated prisoner's dilemma allows the same players to repeat this game over several rounds. If the number of rounds is finite and known in advance by both players (as it is in the current study), the rational prediction is that players should always defect.

In reality, players cooperate at far higher rate than rational predictions, and thereby earn a higher payout than the rational decision-maker (Camerer, 2003). Emotion is evoked to explain these discrepancies but is operationalized quite differently across different threads of research. Figure 1 helps illustrate the different pathways that have been studied, typically in isolation. Our novel analysis focuses on paths C and D, but our results hold indirect implications for paths A and B.

**2.1 Feelings Shape Decision (Path A)**

Several theorists explain the discrepancy between rational and actual behavior by emphasizing the role of internal feelings (Frank, 1988; Lerner, Li, Valdesolo, & Kassam, 2015). For example, Fehr and Schmid (Fehr & Schmidt, 1999) argue that people tend to experience guilt when they exploit their opponent (or contemplate such an action) and anger/envy when they are exploited. Although models differ in detail, a common approach is to fold these emotions into the utility function used to guide decisions. For example, Fehr and Schmidt argue that the pleasure someone feels in an outcome is a mixture of selfish emotion (i.e., related to the individual reward they receive) and "other-directed" emotion (i.e., related to the absolute or relative reward received by others). When people receive more than their opponent, their self-interested pleasure may be reduced by guilt. These models allow that individuals or cultures may differ in the emotions they feel. For example, social value orientation measures if an individual is selfish (i.e., focused on winning) or selfless (i.e., focused on the collective) (Murphy et al., 2011).

It should be noted that this research tends to focus on "one shot" games (like the ultimatum game) as these afford simpler predictions than iterated games. Many studies also focus on emotions unrelated to the task, often called "exogenous emotion" might shape behavior (Path E). For example, participants might watch a disgusting video before playing a game (Lerner, Small, & Loewenstein, 2004). Here, our focus is on "endogenous emotion" – i.e., emotion that arises from the performance of the game.

**2.2 Partner Expressions Shape Actions (Path B)**

Other research minimizes the role of felt emotion in decision-making but rather argues that expressions serve as a communicative tool for shaping expectations of cooperation (Crivelli & Fridlund, 2018). In our own work, we have shown that people form expectations about whether their partner will collaborate or compete based on their pattern of facial displays, and this shapes decisions through a mechanism we call *reverse appraisal* (de Melo, Carnevale, Read, et al., 2014; de Melo & Gratch, 2019). For example, if a partner smiles after mutual-cooperation, this suggests they hold selfless intentions and will likely cooperate again in the future. In contrast, if a partner smiles after exploiting you, this suggests they have selfish intentions and will likely exploit you again in the future.

Reverse appraisal rests on the foundation of appraisal theories of emotions (Lazarus, 1991; Scherer, Schorr, & Johnstone, 2001). Appraisal theories argue that felt-emotion, and associated external expressions, arise from a cognitive assessment of how a specific event relates to the individual's goals. For example, if an individual values cooperation, they will view mutual-cooperation as congruent with this goal, and thus exhibit positive emotions. In contrast, if an individual is purely self-interested (and myopic), they would appraise mutual-cooperation as incongruent with their goals (as a missed opportunity to exploit their partner and earn more money). Reverse appraisal argues that observers of these expressions essentially work backward from how people emotionally react to specific event to recover their goals.

The reverse appraisal view highlights that internal emotional feelings are not strictly necessary to explain departures from rational predictions. Decision-makers can be seen as acting in a cold and rational manner: they are simply utilizing expectations about partner behavior willingness to collaborate, as revealed by their expressions. Though this work does not preclude a role for felt emotion. Indeed, some findings suggest that both felt emotion and partner emotion could combine to shape suggestions (van Kleef, 2008). In the present paper, we examine the potential importance of this affective path.

**2.3 Partner Expressions Shape Feelings (Path C)**

Theories of emotional contagion or mimicry argue that people catch the feelings of those around them (Hatfield et al., 1994; Tsai, Bowring, Marsella, & Tambe, 2013). For example, if my partner smiles, I will feel more joy. Some research suggests this effect is driven by mimicry: people tend to mimic smiles and this increases felt joy through facial feedback (Niedenthal, Mermillod, Maringer, & Hess, 2010) and this has sometimes been argued as a mechanism that underlies the emergence of empathy. Such theories would suggest a clear correlation between internal feelings with partner expressions. The present paper suggests such contagion does not necessarily play a role in player feelings.

Other research argues that contagion and mimicry can reverse depending on the nature of the situation or task. For example, Lanzetta and Englis (Lanzetta & Englis, 1989) found that framing a situation as collaborative (i.e., a winning payout for one side also results in a winning payout for the other) yields contagion, but framing a situation as competition (i.e., a winning payout for one side yields a loss for the other side), creates a "counter-empathetic" response, meaning that partner smiles evoke negative feelings. As the prisoner's dilemma includes both collaborative outcomes (i.e., mutual-cooperation) and competitive outcomes (i.e., exploitation), it affords opportunities for both collaboration and competition, making it unclear if empathy or counter empathy might emerge in this game. In the present paper, our results suggest that partner expressions might help to frame the situation as one involving cooperation or competition.

**2.4 Partner Actions Shape Feelings (Path D)**

Finally, some research has focused on how partner actions shape the player self-reported feelings. This work has also tended to emphasize the connection between feelings and physiological responses. For example, people report anger following unfair offers in the ultimatum game, but also show increased heart-rate deceleration (Osumi & Ohira, 2009) or insula activation (Grecucci, Giorgetta, van't Wout, Bonini, & Sanfey, 2012). This work has also emphasized that these feelings connect with path (A) and shape subsequent behavior. For example, people are more likely to reject unfair offers if they evoke feelings of anger. In the present paper, our results suggest this pathway can be weaker than partner expressions (C).

**2.5 Unit of Analysis**

One of the challenges of studying emotion is selecting the appropriate unit of analysis. Most research on repeated games, such as the prisoner's dilemma, analyze behavior at the level of the entire game (e.g., average cooperation rate, or feelings after the game is complete). This simplifies issues of experimental control as a manipulation (such as inducing mood or altering the partner decision-making policy) can be treated, statistically, as a pure independent variable.

Yet, emotions have been argued to be brief, relatively intense responses caused by a specific event -- in contrast to mood which is longer term and the cause often unclear (Beedie, Terry, & Lane, 2005). In a repeated economic game such as prisoner's dilemma, arguably the most emotionally salient events are the moments when the joint decision is revealed after each round. Indeed, an analysis of facial expressions in a face-to-face prisoner's dilemma finds that people show the most intense facial expressions in the seven seconds immediately following the revelation of each round's joint outcome (Lei, Stefanov, & Gratch, 2020). This suggests that emotional-feelings may be best explained by analyzing the properties of this immediately-preceding event. Unfortunately, the properties of these events could be influenced by the entire history of the game up to that point (e.g., the prior sequence of game-decisions and the pattern of expressions shown by each player). Ignoring this history means we may overlook important factors that shape behavior or feelings on a particular round.

In the present paper, we choose to focus on joint outcomes as our unit of analysis, but characterize these outcomes by the expression and decision policy that preceded their occurrence (e.g., was their partner acting selfishly or selflessly up to this point). We this simplified view of history may obscure important factors that shape feelings and actions and will return to this point in the discussion.

## 3. Prisoner's Dilemma Corpus

To examine the factors that shape players' momentary feelings in the iterated prisoner's dilemma, we analyze an existing corpus of data we have previously collected in our lab. This data was collected as part of a previously published experiment (de Melo & Terada, 2020). In the original study, the focus was on the (B) link of Fig. 1 (how partner expression influences player decisions) and considered the entire game as a unit of analysis (e.g., cooperation-rate over the entire series of rounds). Here we focus on the (C) and (D) links (how partner expressions and actions impact player feelings) and shift the unit of analysis to feelings that arise from

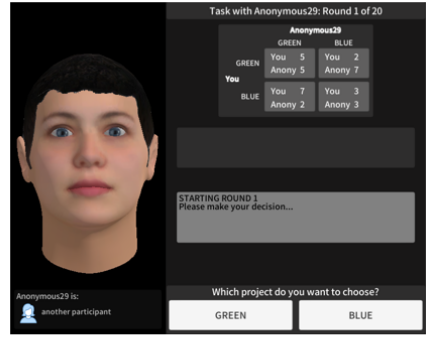

**A) Game Interface**

**B) Agent action policy**

| $p_{CC}$ | $p_{CD}$ | $p_{DC}$ | $p_{DD}$ |
|---|---|---|---|

Cooperation after participant cooperates and counterpart defects

| **Extortion** | | | Round 1: D |
|---|---|---|---|
| 69.2% | 53.8% | 0 | 0 |

| **Generosity** | | | Round 1: C |
|---|---|---|---|
| 100% | 100% | 18.2% | 36.4% |

**C) Agent expression policy**

**Cooperative**

|  | | Counterpart | |
|---|---|---|---|
|  | | Cooperation | Defection |
| Participant | Cooperation | Joy | Regret |
|  | Defection | Anger | Neutral |

**Competitive**

|  | | Counterpart | |
|---|---|---|---|
|  | | Cooperation | Defection |
| Participant | Cooperation | Regret | Joy |
|  | Defection | Anger | Neutral |

**D) Agent Expressions**

Neutral  Joy  Regret  Anger

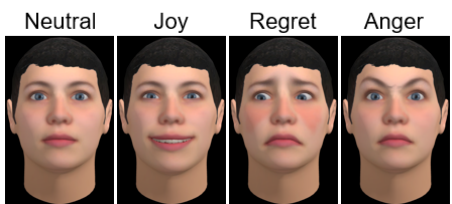

Fig. 2: a) Illustrates the game interface. Players see the payoff matrix and partner expressions; b) shows agent probabilistic action policy; c) illustrates the cooperative versus competitive pattern of partner expressions, which appear as animations that morph from neutral to the poses highlighted in (d). Participants are asked to report their own feelings after learning the joint decision but before seeing their partner's emotional expression.

specific outcomes within the game (e.g., feelings after being exploited by a smiling partner).

Figure 2 summarizes the original experimental design. The prisoner's dilemma task was presented to participants recruited from Amazon Mechanical Turk, framed as an investment game to avoid explicitly labeling decisions as cooperative or non-cooperative ("You can invest in one of two projects: project green and project blue"). Our interest is to investigate the influence of emotion expressions on economic decision making. We predict that participants' economically rational decisions would be affected by emotional signals of others. If the two options in the game were labeled as cooperation and defection, it is possible that decisions would be made based on social norms or morality rather than economic rationality. To eliminate this possibility, we recast the prisoner's dilemma as an investment game, as in previous studies (de Melo & Terada, 2020). Players received points each round based on the following payoff: "if you both invest in project green, then each gets 5 points; if you choose project green but the other player chooses project blue, then you get 2 and the other player gets 7 points; if you choose project blue and the other player chooses project green, then you get 7 and the other player gets 2 points; if you both choose project blue, in which case both get 3 points". Thus, green was the cooperative choice. Each point was converted into a lottery ticket and these were entered into a $30USD lottery. This performance-contingent payoff was in addition to the $2.50 for 20 minutes fixed payment for participation.

Participants believed they were playing another partner online and believed they could see how their partner felt after each round in the game. In reality, participants played computer agents that followed an algorithm that determined the agent's actions and "feelings" as a function of the participants' prior decisions. The agent cooperated or defected based following one of two previously-proposed probabilistic strategies ("extortionist" or "generosity") design to maximize either selfish or collective reward (Press & Dyson, 2012; Stewart & Plotkin, 2013). Expressively, agents displayed "competitive" or "cooperative" pattern of emotional reactions to game events, following an "expression policy" previously shown to convey selfish or selfless intentions (de Melo, Carnevale, Read, et al., 2014). Both the decision and expression policy are contingent, in the sense that the decision or expression differs as a function of the participant's own choices in the game. This design was approved by USC's IRB and participants were briefed of the deception after the study was complete.

Three-hundred and nineteen participants (61% male) were recruited on Amazon Mechanical Turk and were randomized across a 2 × 2 between-participants factorial design: strategy (extortion vs. generosity) × emotion (cooperative vs. competitive). The generous and extortion strategies defined probabilities of cooperation following each of the possible outcomes as shown in Figure 2b. For instance, extortionists never cooperated following a defection by the counterpart, but generous others cooperated following mutual defection with a 36.4% chance. The emotion expressions defined a cooperative and competitive pattern. For instance, cooperative others showed regret following exploitation, whereas competitive others showed joy. The expressions were validated prior to this data collection.

The agent selection probabilities were calculated using Zero-Determinant (ZD) strategies which are memory-one strategies in which the decision for the current round only depends on the outcome of the previous round and they enforce a linear relationship between the players' payoffs in the prisoner's dilemma (Press & Dyson, 2012). ZD strategies are written as a 5-tuple $(p_0, p_R, p_S, p_T, p_P)$, where $p_0$ is the player's probability of cooperation in the first round ($m = 1$), $p_i$ is the probability of cooperation in round $m \geq 2$ given the payoff $i \in \{R, S, T, P\}$ in the previous round. Payoff $R$ and $S$ are given to both players when both player cooperate and defect, respectively. If one player cooperates and the other defects, $T$ is given to the defector and $S$ is given to the cooperator. The relation $T > R > P > S$ is typically assumed to hold. According to Hilbe et al. (Hilbe, Röhl, & Milinski, 2014), the probabilities of cooperation are defined as follows:

$$p_R = 1 - \phi(1 - s)(R - l) \quad (1)$$
$$p_S = 1 - \phi[(1 - s)(S - l) + T - S] \quad (2)$$
$$p_T = \phi[(1 - s)(l - T) + T - S] \quad (3)$$
$$p_P = \phi(1 - s)(l - P) \quad (4)$$

, where $l$, $s$, and $\phi$ are constants.

While ZD strategies are able to enforce a linear relationship between average payoff $\pi$ of the ZD strategist and the expected payoff $\tilde{\pi}$ of the counterpart when the game is repeatedly and infinitely played, Hilbe et al. showed that when the game is played $M$ rounds, the relationship between $\pi$ and $\tilde{\pi}$ follows these inequalities:

$$-\frac{p_0}{\phi M} \leq (1 - s)l + s\pi - \tilde{\pi} \leq \frac{1 - p_0}{\phi M} \quad (5)$$

.

We used the payoff values $T = 7$, $R = 5$, $P = 3$, $S = 2$, and a total number of rounds $M = 20$. The following are the values in our experiment for the constants in Equations (1)-(4), and the relation between $\pi$ and $\tilde{\pi}$ predicted by the inequalities in (5):

Extortion

$$l = P, s = 1/3, \phi = 3/13$$
$$p_0 = 0.000, p_R = 0.692, p_S = 0.000, p_T = 0.538, p_P = 0.000$$
$$\frac{1}{3} \cdot \pi + \frac{2}{3} \cdot 3 - \frac{13}{60} \leq \tilde{\pi} \leq \frac{1}{2} \cdot \pi + \frac{2}{3} \cdot 3$$

Generosity

$$l = R, s = 1/3, \phi = 3/11$$
$$p_0 = 1.000, p_R = 1.000, p_S = 0.182, p_T = 1.000, p_P = 0.364$$
$$\frac{1}{3} \cdot \pi + \frac{2}{3} \cdot 5 \leq \tilde{\pi} \leq \frac{1}{2} \cdot \pi + \frac{2}{3} \cdot 5 + \frac{11}{60}$$

We conducted computer simulations to confirm that the strategies used in our experiment met the zero-determinant requirements (for more detail see de Melo & Terada, 2020).

Findings from the original study indicated that partner expressions and actions interact to determine player action. Participants cooperated more with generous than extortionist others (a main effect of strategy) and more with cooperative than competitive others (a main effect of emotion). Moreover, there was a strategy by expressed-emotion interaction indicating that participants ignored the partners expressed emotions when engaging with extortionists. Overall, the findings emphasize the importance of others' behaviors in promoting cooperation and that, when the strategy is less clear (e.g., tit-for-tat or generosity), people will pay special attention to emotion expressions when deciding whether to cooperate.

The prior analysis, however, did not focus on participants' feelings and what causes them. That is the focus of this paper. To measure participants' emotions, after every round, we asked: "How do you feel about this outcome?" Participants were able to choose among five options: neutral, joy, anger, regret, and anger. Participants also were told that these emotions would be reflected on their avatar, similarly to their counterpart (we will return to the possible impact of this instruction in the discussion). In this new analysis of the data, we look at the relationship between participant's feelings and other's actions and expressions.

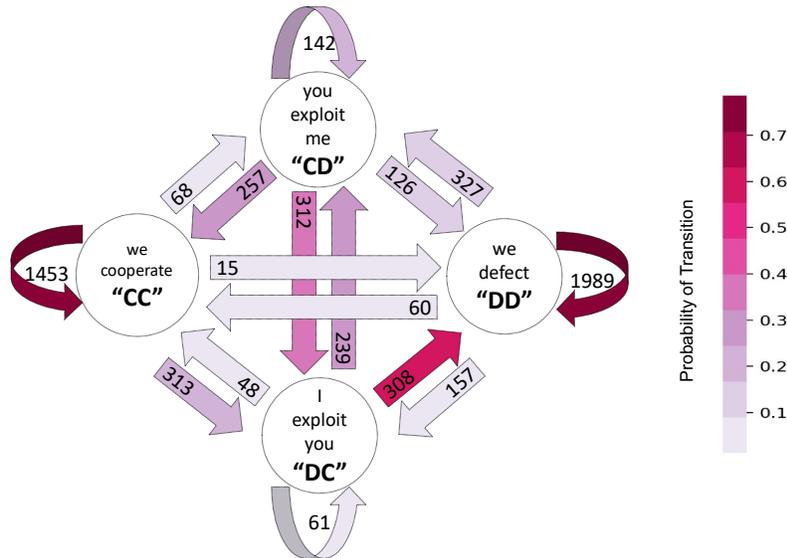

Fig. 3: Illustrates the structure of events within the entire IPD corpus. Circles indicate the four possible joint outcomes and arcs indict how participants reached this outcome (e.g., participants that mutually-cooperate are mostly likely to cooperate again.

## 4. Event-level Analysis of Feelings and Decisions

To analyze feelings and decisions at the level of individual events (i.e., reactions to the joint outcome of a given round), we constructed a database of 6380 individual joint outcomes derived from the 319 participants in the corpus (i.e., 20 outcomes per participant). Fig. 3 provides one view into the structure of these events. For example, most people tended to fall into one of two equilibria (cycling within mutual-cooperation or cycling with mutual-defection) and participants are more likely to move to mutual defection after being exploited. These patterns are consistent with most behavioral studies of the prisoner's dilemma. These events are further organized by experimental condition (e.g., in approximately one quarter of these events, the partner was making decisions according to an extortionist policy but showing a cooperative pattern of expressions).

*A. Feelings after each round*

We examined the participants' self-reported feelings following each round of the prisoner's dilemma across the four conditions. Fig. 4 shows the percentages of participant emotional-feelings reported after rounds of CC (mutual-cooperation), CD (participant cooperation and partner defection), DC (partner defection and participant cooperation), DD (mutual defection) across the four experimental conditions. Taking the experimental factors and round outcome into consideration, joy was the most common emotion experienced (42% on average), but the frequency of joy and other emotions varied considerably by outcome and by experimental factor.

We analyzed the distribution of feelings using a log-linear analysis. This analysis reveals a significant interaction between agent condition (agent strategy and agent expression) and round outcome for participant emotional expression ($G^2=5034.74$, $p<.0001$). The analysis further suggests that partner emotional expression, in combination with round outcome, has the strongest impact on participants' feelings ($G^2$=988.88, p<.0001).

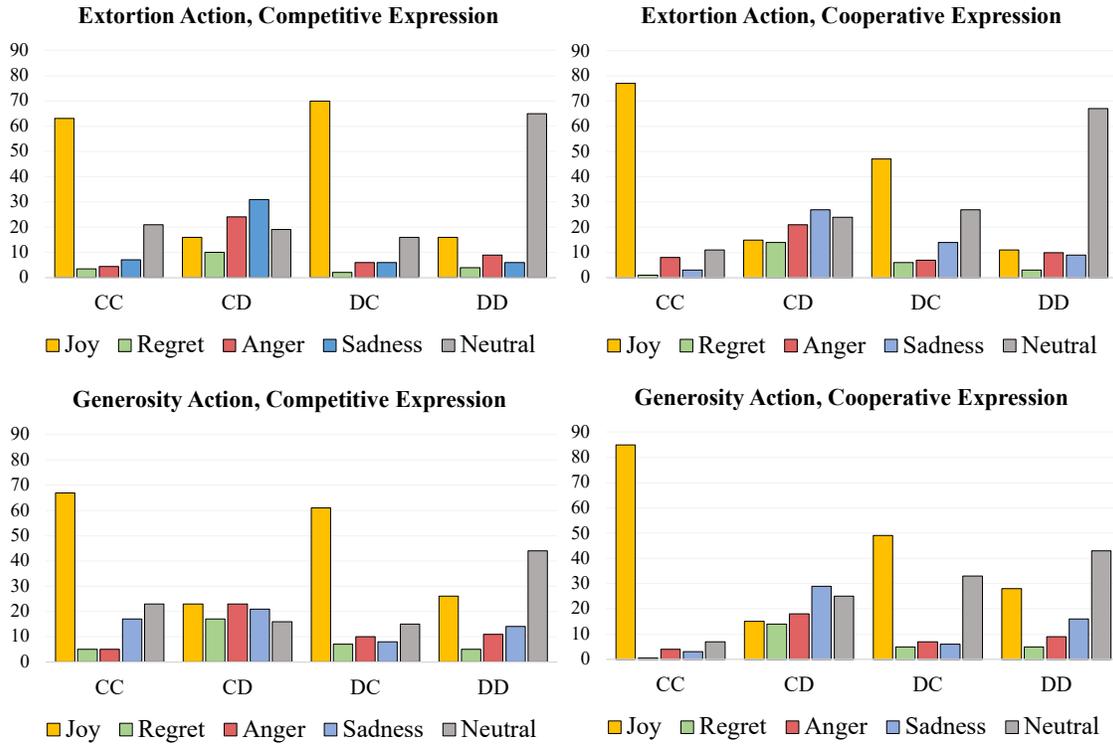

Fig. 4: Distribution of participant-reported emotional-feelings (Joy, Regret, Anger, Sadness, Neutral) following rounds of CC (mutual-cooperation), CD (participant cooperation and partner defection), DC (partner defection and participant cooperation), DD (mutual defection) across the four conditions extortion-competitive, extortion-cooperative, generosity-competitive, generosity-cooperative.

### 4.1 Selfless Feelings

Fig. 4, illustrates that experimental factors have a large impact on feelings of joy following mutual-cooperation and exploitation. In particular, participants facing partners with cooperative expressions experienced more joy-at mutual-cooperation (CC), reflecting thus a selfless pattern of emotion feelings, while participants facing partners with competitive expressions experienced more joy-at exploiting their opponent (DC), reflecting more selfish feelings.

To explore this further, we created a new variable, *selfless feelings,* which indicates the difference in joy-at cooperation versus joy-at competition (i.e., %Joy after CC minus %Joy after DC) – see Fig 5. Conditions with greater selfless feelings indicate that participants feel relatively better about collaborating with than exploiting their partner. Conditions with lower selfless feelings indicate that participants feel relatively better about exploiting them.

There is a significant effect of the experimental conditions on selfless feelings ($G^2$=329.02, p<.0001), and this effect is driven by the contrast between cooperative and competitive partners, which occurs in both the extortion condition ($G^2$=242.22, p<.0001) and the generous strategy condition ($G^2$=232.60, p<.0001). We therefore confirm that participants facing a partner with cooperative expressions report more selfless feelings (e.g., Joy-at mutual collaboration), and those facing competitive partners show less selfless feeling (e.g., Joy-at exploiting their partner).

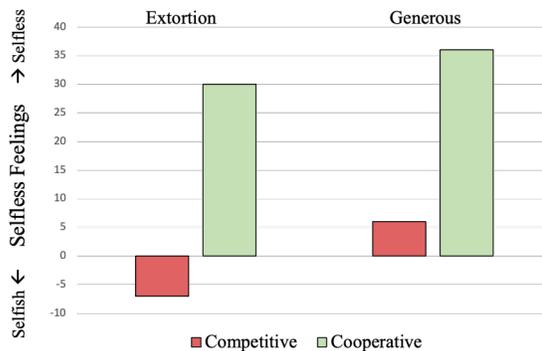 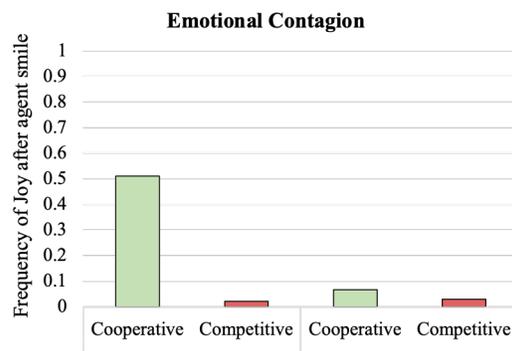

Fig. 5 (Selfless feelings): Shows the extent to which participants reported selfless or selfish feelings as a result of experimental condition. Higher values indicate more joy-at mutual-cooperation (CC) compared with joy-at exploitation (DC).

Fig. 6: Percentage of participants who smiled in round *n+1* following a partner's smile in round *n* across the four conditions (extortion-competitive, extortion-cooperative, generosity - competitive, and generosity-cooperative)

### 4.2 Alternative Mechanisms (Contagion)

The previous analysis suggests that the pattern of agent expressions determines participant feelings, however it is possible that simple emotional contagion could also explain this pattern. To rule this out, we examined if people feel more joy when the agent expressed joy on the previous round. Results are illustrated in Fig. 6.

A log-linear analysis reveals a significant interaction between agent emotion expression and partner strategy on participant joy ($G^2$=1736.32, p<.0001). There are two significant main effects as well: one for partner emotion expression, the other for partner strategy ($G^2$=965.02, p<.0001 and $G^2$=582.28, p<.0001, respectively), but they are almost entirely explained by the interaction effect. As shown in Fig. 4, 51% of participants facing generous-cooperative partners reported joy in round *n+1* following an agent smile in round *n* and only 0.02%, 0.07%, and 0.03% of participants facing generous-competitive, extortionist-cooperative, extortionist-competitive counterparts, respectively, reported emotions consistent with contagion. These results suggest that the difference in participant emotional reactions by partner emotion expression cannot be explained by emotional contagion.

### 4.3 Relationship Between Expression and Action

Up to this point, we have considered how partner actions and expressions influence participant feelings. But do these induced feelings shape player subsequent action? For example, if someone feels joy-at mutual-cooperation, will they be more likely to cooperate again on the subsequent round? Most research on the behavioral impact of feelings have focused on simpler "one-shot" games where players make a single decision. This is because predictions become less obvious in iterated games.

To illustrate the complexity of iterated games, consider the case of joy-at exploiting my partner (DC). A selfish player might experience great joy at successfully exploiting their partner, but how should they act on the next round to achieve this pleasure again. If the selfish player defects again, they might successfully exploit their partner a second time, but they will likely push the partner towards mutual defection for the remainder of the game. If they cooperate, they might encourage their partner back into cooperation (setting up another opportunity to exploit) but they will likely be exploited back by their partner.

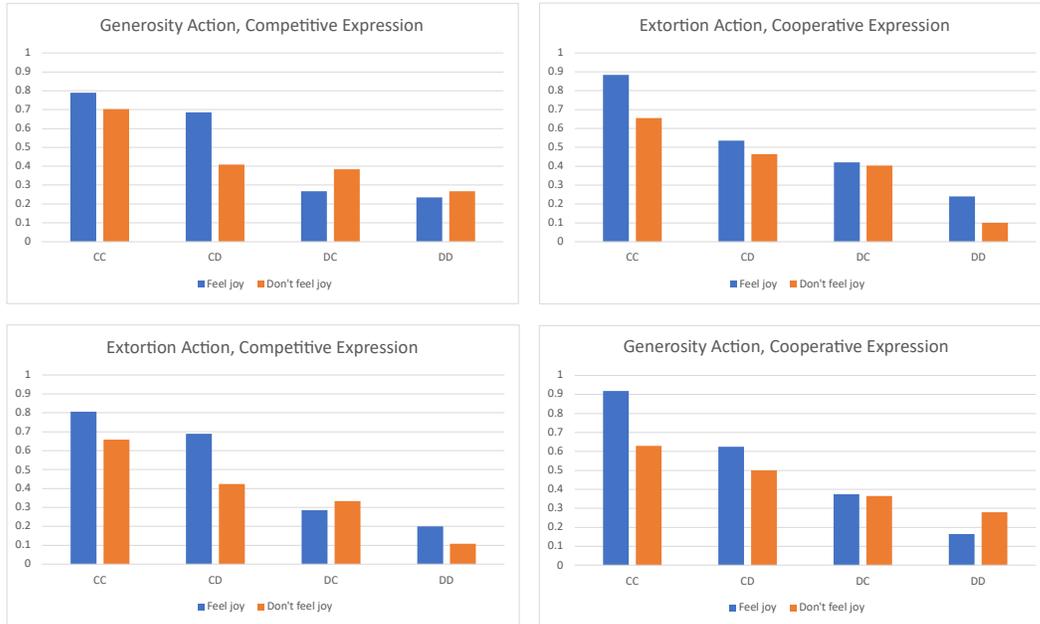

Fig. 7: Frequency of cooperation on round *n+1* after reporting feelings of Joy (or not) as a function of round outcome and condition

The case seems simpler for mutual-cooperation (CC). If a player feels joy-at mutual-cooperation (i.e., a selfless emotion), they should be motivated to cooperate again. If they fail to feel joy-at this outcome (i.e., they feel selfish emotions), they may be motivated to defect.

To explore these patterns, we calculate the rate of cooperation on round *n+1* if the participant experienced a particular outcome on round *n*. To simplify the visualization of this pattern, we distinguish joy from non-joy (i.e., did the participant feel anger, sadness, regret, or neutral on that round). Fig. 7 shows cooperation rates as a function of experimental condition.

While there is a significant interaction between all of these factors ($G^2$=6052.66, p<.0001), an inspection of the percentages of cooperation when they reported joy or not fail to provide clear evidence that feeling alone shapes cooperation. Indeed, the outcome on round *n* seems a better predictor than either feeling or experimental condition. Specifically, people cooperate the most after mutual-cooperation and the least after mutual defection. Whether or not they reported feelings of joy is not especially diagnostic of next action, except for joy-at mutual-cooperation and joy-at being exploited (note that reports of feeling joy-at being exploited and joy-at mutual defection were quite rare).

If we focus on just feelings of joy-at mutual-cooperation, we see that partner expressions have a larger impact than partner actions. There is a significant effect of the partner strategy/action by partner emotion interaction on user choice to cooperate (G2 = 46.20, p < .0001). However, inspection of the main effects reveals that partner expression significantly affects user cooperation (G2 = 42.00, p < .0001), whereas partner action/strategy does not (G2 = 0.58, p = .44). In other words, cooperation rate goes up when partners express joy-at mutual-cooperation compared with those that express joy-at exploitation. In contrast, cooperation rates remain unchanged by partner actions.

A similar pattern can be seen for being exploited (CD), but the raw numbers/frequencies are too low to reach significance. After CD, there is a trend for an effect of the partner strategy/action by partner emotion interaction on user choice to cooperate ($G2 = 6.88$, $p = .14$). Again, inspection of the main effects reveals that partner expression tends to affect user cooperation ($G2 = 2.38$, $p = .12$), whereas partner action/strategy does not ($G2 = 0.38$, $p = .54$).

Overall, the results suggest a weak association between participant feelings and the actions that immediately follow these feelings but there is some evidence that the association between feelings and immediate cooperation is more strongly shaped by partner expression than partner actions in the game.

## 5. Discussion and Limitations

In this paper, we focused on how partner actions and expressions impact a player's moment-to-moment felt-emotions in the iterated prisoner's dilemma game. Prior work focused on the role of partner actions on feelings (e.g., anger-at being exploited) or argued people could "catch" their partner's emotions through emotional contagion. We found that momentary feelings were primarily shaped by the prior pattern of partner emotional expressions (i.e., competitive vs. collaborative), more than their pattern of actions (i.e., extortion vs. generosity). In particular, when a player's partner had previously shown joy-at mutual-cooperation, players reported more joy when they achieved mutual-cooperation. In contrast, when a player's partner had previously expressed joy-at exploiting them, players reported relatively greater feelings of joy-at exploiting them back. Surprisingly, whether or not the player was generally exploitative or generous played far less a role in shaping feelings. The results, therefore, suggest that people are able to discern, consciously or not, the implications of the partner's actions – e.g., what should I do next? – from the partner's emotions – e.g., what is the status of our relationship? In essence, a generous partner can make a player feel worse if the emotions are not congruent with the actions and, in contrast, an exploitative partner can make a player feel better overall by showing selfless emotional expressions. Therefore, whereas the partner's actions may reflect shorter-term consequences in the outcome of the game, the partner's expressions seem to reflect longer-term consequences in the subjective impressions formed about the partner and perhaps the prospects of future interaction.

Our findings connect with and serve to extend Lanzetta and Englis's research on counter-empathy (Lanzetta & Englis, 1989) – Fig. 1, path (C). They found that people experienced counter-empathic feelings when instructed a situation was competitive (versus cooperative), or that the task had a win-lose payout (versus win-win). Here we did not instruct people on the nature of the task and prisoner's dilemma involves a mixture of win-win and win-lose incentives (and players are free to self-select cooperation or competition). Yet people developed empathetic or counter empathic feelings based on their partner's expressions. It is possible that the partner's pattern of emotional expressions served to define the nature of the situation. It is as-if participants were told they are in a competitive situation if their partner showed competitive expressions. Strikingly, this was not signaled by if the partner *actions* were completive or collaborative.

These findings also have implications for reverse appraisal methods models of social decision-making (e.g., de Melo, Carnevale, Read, et al., 2014). These theories tend to focus on the link between partner expression and player action – Fig. 1, path (A) – but our findings highlight the potential mediating role of felt emotion. More studies are required to explore the importance of such feelings in predicting behavior.

There are a number of important limitations and qualifications to these results. One chief concern is to whether self-reported feelings truly reflect a participant's experience of emotion. Although we instructed participants to report "How do you feel about this outcome?", participants also knew their answers were visible to their partner. As a result, there is the possibility that this motved them to either regulate their own emotions or alter their answers to report more socially desirable answers. Other research has shown that people regulate their facial expressions in the prisoner's dilemma when they know they are being watched (Hoegen, Gratch, Parkinson, & Shore, 2019).

The possibility of emotion regulation does not make these results less interesting but does change their interpretation. When faced with a partner that freely shows joy-at exploiting them, this may signal that "impolite" expressions are considered acceptable. Thus, players may more authentically report their true feelings of schadenfreude. Further studies are needed to disambiguate these processes along the lines of Shore and Parkinson (Shore & Parkinson, 2018).

There are some patterns in the results that deserve further analysis. Here we largely focused on feelings of joy-at cooperation versus joy-at exploitation, but Fig. 3 also shows differences in when people reported no emotion (i.e., neutral). People report more neutral after mutual-defection with the extortionist agent. Though it should also be noted that mutual-defection is the most frequent state with this agent, so this could simply reflect attenuation of feelings when locked into an unproductive pattern.

This paper highlights a number of important methodological advancements but these techniques introduce some caution in how to interpret the findings. One distinction is between "emotion" and "emotion-at". Most emotion recognition research asks third party observers to guess how individuals feel (or sometimes asks participants to self-report their feelings). Here we ask people to report how they feel about a specific referent – i.e., emotion-at. More research is needed to understand how this change might alter results. Perhaps responses might differ if the referent was not made salient. Third-party ratings might also differ depending on if the context is visible to the rater.

Another innovation was to focus on moment-to-moment feelings rather than emotions over the entire interaction. However, this introduces a quasi-experimental design because, at the level of individual rounds, participants are selecting their own outcomes (i.e., whether or not a player experienced exploitation depends on their own choices). That means we cannot firmly establish causality. Nonetheless, we argue such analyses are important if one wishes to study interaction. More fully-controlled designs sacrifice true interactivity in an attempt to maintain experimental control. Thus, multiple lines of evidence will be needed to fully establish the phenomena we report.

Here we focused narrowly on the question of how partner expressions and actions influenced feelings, but these results need to be reconciled and integrated with existing findings on the other paths illustrated in Fig. 1. For example, reverse appraisal research argues that expressions serve as information about the partner and one does not need to posit emotion to explain changes in player actions. Some frameworks seek to integrate different pathways. For example, Van Kleef's Emotion as Social Information (EASI) framework allows both an informational and evocative pathway for partner expressions to influence the player (van Kleef, 2008). Far more empirical work is needed to assess the utility of this model.

Finally, we draw attention to broader ethical considerations when designing cognitive systems that use affective signals. Some have argued that any use of emotional expressions by a machine is unethical because machines don't have emotions and therefore any expression is, by definition deceptive (Bringsjord & Clark, 2012). These arguments are naïve in that they assume that human expressions of emotion always truthfully reflect underlying feelings, which is clearly not the case.

Rather, expressions frequently serve as communicative tools that help guide and shape and facilitate social interactions (e.g., (Crivelli & Fridlund, 2018)). For example, an expression of regret can signal an honest intention to repair a relationship after a social transgression and can promote prosocial ends, even if the person (or machine) doesn't feel actual regret (Rychlowska, van der Schalk, Gratch, Breitinger, & Manstead, 2019). Such regulated signals, though not reflecting genuine emotion, serve important social functions. In the extreme case, some even argued that explicit deception is acceptable, even desirable, if the deception benefits the target of the lie (Levine & Schweitzer, 2015). The results presented here demonstrate the potential for cognitive systems to use emotion to serve this multitude of social functions. We show that agents can successfully facilitate cooperation by aligning their expressions with an underlying intention to cooperate (i.e., the cooperative generous condition). We also show that agents can use emotion to mitigate the consequences of competitive actions it has to take, for reasons that may be beyond its control (i.e., the cooperative extortion condition). The potential ethical complication is that cognitive systems have the ability to manipulate human feelings in anti-social ways even when these expressions are entirely divorced from the agents underlying intentions. This effect persists even in repeated interactions. This highlights the potential power of these expressions to shape feelings in both prosocial and selfish ways. It also highlights an important distinction between a focus on deception versus a focus on intent. Proscriptions that ban affective signals from machines with the argument that they are deceptive will eliminate an important coordination tool. Rather, we suggest ethicists focus on the social function and intent of synthetic expressions, and designers on the opportunity of using emotion in cognitive systems to build a more cooperative society.

## Acknowledgements

Research as sponsored by the Army Research Office and was accomplished under Cooperative Agreement Number W911NF-20-2-0053. The views and conclusions contained in this document are those of the authors and should not be interpreted as representing the official policies, either expressed or implied, of the Army Research Office or the U.S. Government. The U.S. Government is authorized to reproduce and distribute reprints for Government purposes notwithstanding any copyright notation herein.